\documentstyle[preprint,epsf,aps]{revtex}
\tightenlines
\begin{document}
\title{Influence of damping on the vanishing of the electro-optic effect in chiral isotropic media}

\author{G.S. Agarwal$^{1}$ and Robert W. Boyd$^{2}$}
\address{$^{1}$Physical Research Laboratory, Navrangpura, Ahmedabad-380
009, INDIA\\
$^{2}$Institute of Optics, University of Rochester, Rochester, NY14627, USA}
\date{\today}
\maketitle
\begin{abstract}
Using first principles, it is demonstrated that radiative damping alone cannot
lead to a nonvanishing electro-optic effect in a chiral isotropic medium. This
conclusion is in contrast with that obtained by a calculation in which damping
effects are included using the standard phenomenological model.  We show that
these predictions differ because the phenomenological damping equations are
valid only in regions where the frequencies of the applied electromagnetic
fields are nearly resonant with the atomic transitions.  We also show that 
collisional damping can lead to a nonvanishing electrooptic effect, but with a
strength sufficiently weak that it is unlikely to be observable under realistic
laboratory conditions.

\end{abstract}
\newpage
Several recent papers \cite{buck,stedman} have discussed the question of
properly taking into account
various relaxation processes while calculating the nonlinear response of an
optical system. Even the existence of certain nonlinear optical processes is
thought to be closely linked to the existence of a damping mechanism
\cite{silberberg,kauranen,verbiest}. In this connection, it is especially
important to incorporate in a consistent manner the effects of relaxation
processes.  Very often the nonlinear response \cite{boyd} is calculated by
modifying the equation for the off-diagonal elements of the
density matrix (coherences) by introducing phenomenological relaxation terms as
follows:
\begin{equation}
\frac{\partial\rho_{ij}}{\partial t} = -i\omega_{ij}\rho_{ij}+ {\rm field~
terms~ \Rightarrow}
\end{equation}
\begin{equation}
\frac{\partial\rho_{ij}}{\partial t} = -i\omega_{ij} \rho_{ij} - \Gamma_{ij}
\rho_{ij} + {\rm different~field~terms}.
\end{equation}
The equations for the populations are also modified appropriately. Such
modifications
have been extensively used in nonlinear optics and even have led to the prediction
of new effects such as collision-induced resonances which have been
subsequently been observved experimentally 
\cite{lotem}.  Kauranen and Persoons \cite{kauranen} have recently presented a
theoretical argument that predicts the existence of a linear electro-optic
effect (EOE) in chiral isotropic media provided material damping is taken
into account. Their
result follows by using (2). However, it is not clear a priori if Eq.~(2) can
be used
to describe the linear electro-optic effect. In order to see the origin of
this uncertainty,
let us examine the expression for the nonlinear susceptibility describing
the electro-optic
effect in a chiral isotropic medium. The derivation given in Ref. \cite{kauranen} is
based on the standard
phenomenological equations (2) which take into account various damping
processes in the
medium. The nonlinear susceptibility is shown to have contributions of the form
\begin{equation}
X
\equiv\frac{2i\gamma_{ng}}{{(\omega_{ng}-i\gamma_{ng})}{(\omega_{ng}+i\gamma_{ng
})}
{(\omega_{mg}+\omega+i\gamma_{mg})}}~.
\end{equation}
The authors of ref. \cite{kauranen} have suggested that this damping-dependent
contribution is
the one which can lead to a nonvanishing electro-optic effect in a chiral
isotropic medium. Let us examine this contribution further. We note first that
the usual expression for the second-order susceptibility consists of two
energy denominators whereas the above contribution consists of three. Clearly
such a term arises from the combination of two contributions as
$X$ can be written as
\begin{equation}
X =
\frac{1}{(\omega_{mg}+\omega+i\gamma_{mg})}\Bigg[\frac{1}{\omega_{ng}-i\gamma_{
ng}} - \frac{1}{\omega_{ng}+i\gamma_{ng}}\Bigg]~.
\end{equation}
We note also that denominators such as $(\omega_{ng}-i\gamma_{ng})$ do not have an
optical frequency contribution. Such denominators arise from the interaction of
the system with a zero-frequency field. We show below that in a correct treatment of
radiative damping,
the denominator should be replaced by ones that involve 
frequency-dependent damping coefficients. Thus a first-principles treatment 
would lead to
\begin{equation}
X\equiv\frac{1}{(\omega_{mg}+\omega+i\gamma_{mg}(\omega))}\Bigg[\frac{1}{(\omega
_{ng}
-i\gamma_{ng}(0))}-\frac{1}{(\omega_{ng}+i\gamma_{ng}(0))}\Bigg].
\end{equation}
Note that the frequency dependence of $\gamma$ in each denominator depends on
the frequency component of the electromagnetic field responsible for such a
denominator. Thus the denominators corresponding to the static field have
dampings evaluated at zero frequency. As
discussed below, for the case of radiative damping $\gamma_{ng}(0)$ vanishes
identically, which implies that $X=0$. Thus a
first principles (and correct) treatment of radiative damping does {\it not} lead to
any electro-optic effect in a chiral isotropic medium. We also show below that $X$ is
at most very small for the case of collisional damping.  The nonvanishing of the EOE
effect reported earlier is due to inappropriate use of equations which are not valid
for the calculation of the EOE effect. Thus, when using the modification (2) in the
calculation of the nonlinear optical response, one has to keep in view the conditions
under which (2) has been derived. This need necessitates an examination of the
microscopic  theory leading to the derivation of the result (2). It may also be noted
that,  in recent times, one has discovered a number of other interesting situations 
which cannot be described by equations like (2). For example, there are situations
under which the coherences get coupled to the populations, and this situation has led
to considerable work on quantum interferences
\cite{ficek}. In addition, there is the subject of inhibited spontaneous emission
where the modifications of (2) due to strong  external fields play an important role
\cite{lange}. 

In order to uncover the role of relaxation mechanisms on the 
response to external fields and to determine how relaxation depends on the
frequency of the applied field, we consider first a very simple model. This model 
brings out the salient features of the problem and enables us to establish that
the form of the damping operator depends on the various frequency scales in the
system.  We consider the case in which the medium can be described by a
one-dimensional harmonic oscillator with displacement $x$ and with frequency $\omega_o$.
Let the medium interact with an external electromagnetic field of frequency $\omega$
described by 
\begin{equation}
E = {\cal E} e^{-i\omega t} + {\cal E}^* e^{i\omega t}.
\end{equation}
The equation of motion with a phenomenological damping constant $\Gamma$ is
\begin{equation}
\ddot{x}+\Gamma\dot{x}+\omega^2_0 x = \frac{e \cal E}{m} e^{-i\omega t}
+\textrm{ c.c.}~.
\end{equation}
The response of the medium can then be expressed as
\begin{equation}
e~ x(t)= \chi(\omega){\cal E} e^{-i\omega t}+ \textrm{c.c.},
\end{equation}
\begin{equation}
\chi(\omega) = \frac{e^2}{m(\omega_0^2 - \omega^2 - i\omega\Gamma).}
\end{equation}
In this manner one obtains the familiar response function. We would like to
examine
whether the response $\chi(\omega)$ as given by Eq.~(9) is valid for all
frequencies. Thus we would like to understand if the introduction of a
{\it frequency-independent} damping constant $\Gamma$ in Eq.~(7) is justified for all
frequencies of the
applied electromagnetic field. For this purpose we start from first
principles. Let us
consider the interaction of the system oscillator with a bath. The bath will be
responsible for the relaxation processes described phenomenologically by the damping
parameter $\Gamma$ in Eq.~(7). As usual we model the bath by a set of harmonic
oscillators. The Hamiltonian for the system oscillator interacting with a bath is
given by
\begin{equation}
H=\frac{p^2}{2m} + \frac{1}{2} m \omega_0^2~ x^2 - e~x {\cal E} (t) - x F(t),
\end{equation}
where  ${\cal E}$(t) is the time-dependent electromagnetic field and F(t)
represents the effect of the bath terms
\begin{equation}
F(t) = \sum_{j} (g_j a_j e^{-i\omega_j t} + {\textrm h.c.}).
\end{equation}
Here $\omega_j (>0)$ are the frequencies of the bath oscillators $a_j$ and
$g_j$ are
the coupling constants of the system oscillator with the bath oscillators.
The Heisenberg equations can be easily derived from (10):
\begin{eqnarray}
\dot{x} &=& p/m,~~\dot{p} = -m\omega_0^2 x + e{\cal E}(t) + F(t),\nonumber\\
\dot{a}_j &=& i g_j^* e^{i\omega_j t} x(t).
\end{eqnarray}
We integrate formally the equation for $a_j$ and substitute it into the
equation for
$p$ to obtain
\begin{equation}
\dot{p} = -m\omega^2_0 x + e{\cal E}(t) + F_0 (t) +\int_0^t
K(t-\tau)x(\tau)d\tau,
\end{equation}
where
\begin{equation}
F_0 (t)=\sum_{j} g_j a_j (0)e^{-i\omega_j t} + {\textrm h.c.}~,
\end{equation}
\begin{equation}
K(t-\tau) = (i \sum_{j} |g_j|^2 e^{-i\omega_j(t-\tau)} +\textrm{c.c.}).
\end{equation}
Note that Eq.~(13) is derived without any approximation. The further
simplification will depend on the values of $|g_j|,\omega_j, \omega$ etc. Let
us examine the average response for the case in which ${\cal E}(t) = {\cal E} e^{-i\omega t}
+\textrm{c.c.}$
Note that the mean value of the operator $a_j (0)$ is zero  and hence $\langle
F_o(t)\rangle = 0$.
It should be borne in mind that
$\omega$ is positive. Using Eqs.~(13) and (14), taking quantum mechanical expectation values
and the long-time
limit
$t\rightarrow\infty,$ we obtain
\begin{equation}
\langle x \rangle = \frac{e {\cal E} e^{-i\omega t}}{m(\omega_0^2 -
\omega^2) - K(\omega)} + \textrm{c.c.},
\end{equation}
where
\begin{eqnarray}
K(\omega) &=& \lim_{\epsilon\rightarrow 0}\sum_{j} |g_j|^2 i\left\{
\frac{1}{\epsilon+i(\omega_j
-\omega)}- \frac{1}{\epsilon-i(\omega_j + \omega)}\right\}\nonumber\\
&=& K^{'} (\omega)+ i K^{''} (\omega),
\end{eqnarray}
\begin{equation}
K^{''} (\omega) = \sum_{j} |g_j|^2 \pi \delta (\omega_j - \omega).
\end{equation}
The exact result (16) has the same structure as (9) except for the important
difference that $\omega\Gamma$ is replaced by a function $K^{''}(\omega)$
which is dependent
on the frequency $\omega$ of the applied electro-magnetic field. In addition
there is a dispersive contribution \textrm{Re} K$(\omega).$ Note further that
very often one
replaces (18) by
\begin{equation}
K^{''}(\omega)\approx\sum_{j} |g_j|^2 \pi \delta(\omega_j - \omega_0).
\end{equation}
Clearly this can be done if the frequency $\omega_0$ of the system
oscillator is
very close to the applied frequency, i.e., essentially in the resonance
region. If
the frequency $\omega$ happens to be far away from a resonance frequency, then
the phenomenological equation (7) should not be used. This is the real reason
why usage of equations like (2) and (7) can give rise to incorrect nonlinear
optical response for applied frequencies far away from the transition
frequencies. We also find from (18) that for static response
\begin{equation}
\lim_{\omega\rightarrow 0} K^{''} (\omega)\rightarrow \left.\pi|g_{j}
|^2\right|_{\omega_j=0}\rightarrow 0 ,
\end{equation}
for the usual radiative coupling. Thus the static response functions would be
independent of the damping term. More generally no damping term can appear in
the static response as long as the bath does not have a characteristic static
frequency.

The features discussed above are valid rather generally. To see this we
consider the dynamical equations for a two-level system undergoing, say, 
radiative damping. The case of a two-level system is more involved because
of the intrinsic nonlinearity of the two level system. However the salient
features can be uncovered by using the wavefunction approach. Let us write the
interaction Hamiltonian of a two level system interacting with the field and
undergoing radiative damping, as
\begin{eqnarray}
H&=&\hbar\omega_0 |e\rangle \langle e| - \hbar [G(t)|e\rangle \langle g | +
\textrm{h.c.}]
- \hbar\sum_{k} (g_k a_k e^{-i\omega_k t} |e\rangle\langle g | +
\textrm{h.c.}),
\end{eqnarray}
where we sum over all field modes labelled by the index $k$ and where,
\begin{eqnarray}
G(t) &=& \vec{d}.\vec{E}(t)/\hbar. \\ \nonumber
 &=& G_0 e^{-i\omega t} + \textrm{c.c.} ~.
\end{eqnarray}
The last term in (21) is responsible for the radiative decay of the atom. The
coupling to the mode $k$ with frequency $\omega_k$ of the electromagnetic field
is represented by $g_k$; $a_k$ is the photon annihilation operator. The
wavefunction of the whole system can be expressed as
\begin{equation}
|\psi\rangle = \psi_e | e, \{0\}\rangle + \psi_g| g ,\{0\}\rangle + \sum_{k}
\psi_k | g, \{k\}\rangle,
\end{equation}
where $ \{0\} (\{k\})$ represents the vacuum (one photon in mode $k$) state
of the field.
The Schr\"{o}dinger equation leads to
\begin{eqnarray}
\dot{\psi}_g &=& i G^* (t) \psi_e ,\nonumber \\
\dot{\psi}_e &=& - i\omega_0\psi_e + i G(t)\psi_g + i\sum_{k} g_k e^{-i\omega_k
t}\psi_k,\nonumber \\
\dot{\psi}_k &=& i g_k^* {\psi_e} e^{i\omega_k t}.
\end{eqnarray}
The initial conditions are $\psi_e = \psi_k = 0, \psi_g = 1.$ The induced
polarization is to be obtained from the off-diagonal element $\rho_{eg}=\psi_e
\psi_g^*$. Note that to first order in the applied electromagnetic field,
$\rho_{eg}$ is
\begin{eqnarray}
\rho_{eg}^{(1)}(t)&=&
\psi_e^{(1)} (t) \psi_g^{*(0)} (t) + \psi_e^{(0)} (t) \psi_g^{*(1)}
(t)\nonumber\\
&=& \psi_e^{(1)} (t)
\end{eqnarray}
\begin{equation}
\stackrel{t\rightarrow\infty}{\longrightarrow}~~\psi^{(+)} e^{-i\omega t} +
\psi^{(-)} e^{i\omega t} .
\end{equation}
To obtain the steady state response we combine last two equations in (24)
\begin{equation}
\dot{\psi_e}= -i\omega_0 \psi_e + i G(t)\psi_g - \sum_{k}|g_k|^2\int_0^t
e^{-i\omega_k(t-\tau)} \psi_e(\tau)d\tau,
\end{equation}
and thus to first order in the external electromagnetic field we obtain
\begin{equation}
\dot{\psi}_e^{(1)} = -i\omega_0 \psi_e ^{(1)} + i G(t)-\sum_{k}|g_k|^2\int_0^t
e^{-i\omega_k(t-\tau)}\psi_e^{(1)} (\tau)d\tau.
\end{equation}
In terms of Laplace transforms we have the result
\begin{equation}
\hat{\psi}^{(1)}_e (z) = \left\{z+i\omega_0 + \sum_{k}|g_k|^2(z+i\omega_k)^{-1}
\right\}^{-1} i\left\{G_0 (z+i\omega)^{-1} + G_0^* (z-i\omega)^{-1}\right\},
\end{equation}
where we have used the explicit form (22) From (29) we get the response in the
long time limit
\begin{eqnarray}
{\psi}^{(+)} &=& \left(i(\omega_0 - \omega) + \sum_k|g_k|^2(i\omega_k -
i\omega)^{-1}\right)^{-1} G_0, \nonumber \\
{\psi}^{(-)} &=& \left(i(\omega_0 + \omega) + \sum_{k}|g_k|^2 (i\omega_k +
i\omega)
^{-1}\right)^{-1} G_0^*.
\end{eqnarray}
The induced polarization can now be calculated
\begin{eqnarray}
\vec{p}(t) &=& (\rho_{eg} \vec{d}_{eg}^* + \textrm{c.c.})\nonumber\\
&\equiv&\vec{p}_0 e^{-i\omega t} + c.c.~,
\end{eqnarray}
where $\vec{p}_0$ is calculated using Eq.~(30) as
\begin{equation}
\vec{p}_0 = \psi^{(+)} \vec{d}_{eg}^* + \psi_e^{(-)*}\vec{d}_{eg}.
\end{equation}
This is the most general result for the linear response. No assumption has been
made regarding the nature of the bath.
It should be borne in mind that all frequencies in (30) are positive. The
radiative corrections enter the response function through the quantity
\begin{equation}
K (z) = \sum_{k} |g_k|^2 (z+i\omega_k)^{-1}.
\end{equation}
It should be noted that the actual radiative correction terms depend on the
frequencies of the applied fields rather than the atomic frequencies. It is
only
when the applied frequency is close to the atomic frequency that  we can
use the
approximate replacement $\omega\rightarrow\omega_0$ in $\psi^{(+)}$ (this
cannot be done in $\psi^{(-)})$. We thus find that the counter-rotating
contribution $\psi^{(-)*}$ in Eq.~(32) does {\it not} depend on the
radiative damping \cite{louden}. 
The rotating-wave contribution depends on the radiative damping;
however the radiative damping is to be evaluated at the applied frequency. If
such an applied frequency is very far from the atomic transition,  as for
example
for dc fields, then {\it no} radiative damping term appears in the
response. Thus the full quantum mechanical calculation also leads to the same
conclusion as we derived for the simple oscillator model. Further the above
analysis can be easily extended to the multilevel systems and to the
calculation
of second-order and higher-order response. We find similar conclusions
regarding
the various denominators which appear in response functions. The argument given
in the context of Eqs.~(4) and (5) is correct and we rule out the possibility of
the occurence of electro-optics effect due to radiative damping.

A pertinent question could be: can other damping mechanisms, such as phase
changing collisions, possibly lead to the nonvanishing of the EOE in 
isotropic chiral medium? This question has to be examined by considering a detailed 
microscopic model for the collisional process. However, a simple model 
calculation outlined below suggests that even if the effect is
non-vanishing it must be extremely small; in particular it must exponentially small in
a large quantity.

Consider the equation for the optical coherence $\sigma\equiv\rho_{eg}$. 
Let $f(t)$ be a stochestic source which represents the effect of phase 
changing collisions. We model $f(t)$ to be a Gaussian stochastic process 
with correlations given by
\begin{equation}
\langle f(t)\rangle=0,~~~~~ \langle f(t)f(\tau)\rangle=e^{-\Gamma|t-\tau|}~~ f_0^2 .
\end{equation}
Here $\Gamma^{-1}$ is the magnitude of the collision time. The equation for the
optical coherence can be written in the form
\begin{equation}
\dot{\sigma}=-i\Delta\sigma - i f(t)\sigma+iG,
\end{equation}
where $G$ represents the external field. If $\Gamma^{-1}$ is the smallest
time scale in the problem, then one can show using the standard methods
\cite{vankampen} that 
\begin{equation}
\langle\sigma\rangle = {\left(\frac{f_0^2}{\Gamma}+i\Delta\right)}^{-1} (iG).
\end{equation}
Thus one recovers the result of the phenomenological theory. However, for the response
to a static field,  
$\Delta$ is of the order of the optical frequency whereas typical collisional process
take place over a scale which is of the order of psec or larger. Thus $\Gamma^{-1}$
is no longer the smallest time in the problem. The smallest time scale will instead be
$\Delta^{-1}$. In such a case one can show that in the long-time limit
\begin{eqnarray}
\langle \sigma \rangle &=& iG\int_0^{\infty} d\tau e^{-i\Delta\tau} \exp \left\{
-\frac{f_0^2}{\Gamma} \left(\tau - \frac{(1-e^{-\Gamma\tau)}}{\Gamma}\right)\right\}
\nonumber \\ &\approx& iG \int_0^\infty d\tau e^{-i\Delta\tau} e^{-\frac{1}{2}
f_0^2\tau^2}. 
\end{eqnarray}
Note that the square bracket in Eq. (4) is just the real part of 
$\int_{0}^{\infty}d\tau e^{-i\omega_{ng}\tau-\gamma_{ng}\tau}$,  and thus 
if we had treated the damping properly it has to be replaced by
\begin{eqnarray}
[~ ]&\rightarrow& {\rm Real} \int^{\infty}_{0} d\tau e^{-i\omega_{ng}\tau -
\frac{1}{2} f_{ng}^2\tau^2} \nonumber \\
&=& \sqrt\frac{\pi}{2 f^2_{ng}} \exp\left(-\frac{\omega^2_{ng}}{2
f^2_{ng}}\right).
\end{eqnarray}
Thus collisional damping can make the EOE in chiral isotropic medium nonzero.
However it would be extremely small unless the strength of collisions is
comparable to $\omega_{ng}$ i.e., $\omega_{ng}\sim f_{ng}$.

In conclusion, we have shown that radiative damping cannot lead to a nonvanishing EOE
in a chiral isotropic material.  For the case of collisional damping, a nonvanishing
EOE is predicted, but the magnitude of this effect is expected to be so small that
it is unlikely that this effect could be observed experimentally.  These results are
in contrast with recent suggestions that relaxation effects can lead to an EOE in
chiral isotropic materials, with potentially important practical implications.  More
generally, we have shown that in general it is not adequate to use a
frequency-independent damping parameter in treating relaxation processes within the
context of density matrix calculations.

The authors gratefully acknowkledge useful discussions with M. Kauranen and with P. W.
Milonni.  G.~S.~Agarwal thanks the National Science Foundation grant no. INT-9605072
which made this collaboration possible.   R.~W.~Boyd gratefully acknowlwedges 
support  by the Office of Naval Research, the Army Research Office, the
Air Force Office of Scientific Research, and the U.S. Department of Energy Office of
Science.

\end{document}